\documentclass[12pt]{iopart}

\usepackage{graphicx}
\begin{document}

\title[Modeling the formation of the $^{13}$C neutron source in AGB stars]{Modeling the formation of the $^{13}$C neutron source in AGB stars}

\author{D Vescovi$^{1,2,3}$ \& S Cristallo$^{3,2}$}

\address{$^1$ Gran Sasso Science Institute, Viale Francesco Crispi, 7, 67100 L'Aquila, Italy}
\address{$^2$ INFN, Section of Perugia, Via A. Pascoli snc, 06123 Perugia, Italy}
\address{$^3$ INAF, Observatory of Abruzzo, Via Mentore Maggini snc, 64100 Teramo, Italy}
\ead{diego.vescovi@gssi.it}
\vspace{10pt}
\begin{indented}
\item[]November 2019
\end{indented}

\begin{abstract}
A major source of uncertainty in AGB models is the partial-mixing process of hydrogen, required for the formation of the so-called $^{13}$C pocket.
Among the attempts to derive a self-consistent treatment of this physical process, there are 2D and 3D simulations of magnetic buoyancy. 
The $^{13}$C pocket resulting from mixing induced by magnetic buoyancy extends over a region larger than those so far assumed, showing an almost flat $^{13}$C distribution and a negligible amount of $^{14}$N. Recently, it has been proved to be a good candidate to match the records of isotopic abundance ratios of \textit{s}-elements in presolar SiC grains. However, up to date such a magnetic mixing has been applied in post-process calculations only, being never  implemented in a stellar evolutionary code. Here we present new stellar models, performed with the 1-d hydrostatic FUNS evolutionary code, which include magnetic buoyancy. We comment the resulting \textit{s}-process distributions and show preliminary comparisons to spectroscopic observations and pre-solar grains measurements.
\end{abstract}

%
\vspace{2pc}

\section{Introduction}
The existence of about half of the nuclei heavier than iron can be explained through neutron (\textit{n}) captures occurring in the Asymptotic Giant Branch (AGB) phase of Low-Mass Stars (LMS).
In this process, because of the longer timescale, \textit{n}-captures usually occur at a slow (\textit{s}) rate compared to the $\beta$-decay of unstable nuclei. For this reason it is common to refer to it as \textit{s}-process.

While the dynamics of the \textit{s}-process, as well as the major reaction responsible for the supply of neutrons, i.e. the $^{13}{\rm C}(\alpha,{\rm n})^{16}{\rm O}$ reaction, are well-known \cite{cris18}, the mechanism through which $^{13}{\rm C}$ neutron source is made available is still missing.
Classical models typically assumes the formation of a thin layer enriched in $^{13}{\rm C}$, the so-called $^{13}{\rm C}$-pocket, in the He-rich region below the convective envelope of an AGB star, as a conquence of some partial-mixing of envelope hydrogen with the products of helium burning \cite{buss99}.
Recently, many physically-based approaches have been developed in order to model the penetration of proton-rich material from the convective envelope to the He-intershell, involving an opacity-induced overshoot \cite{cris09} or a mixing induced by internal gravity waves \cite{batt16}.
Another approach suggests that the magnetic activity of LMS stars could induce the buoyancy of the material of He-intershell to the envelope \cite{nucc14}. The ensuing mixing would then guarantee, by mass-conservation the necessary downflow of hydrogen-rich material \cite{trip16} for the formation of the $^{13}{\rm C}$-pocket. 
Such $^{13}{\rm C}$-pocket was shown to be able to account both for the solar distributions of \textit{s}-only isotopes \cite{trip16} and for isotopic ratios of \textit{s}-elements measured in pre-solar SiC grains \cite{palm18}.

Here we adopted the formalism developed in \cite{nucc14} for describing the magnetically-driven expansion of material in the radiative He-intershell, derived the corresponding radial velocity of the induced proton downflow, and implemented such magnetic mixing for the formation of the $^{13}{\rm C}$-pocket in the FUNS hydrostatic stellar evolutionary code \cite{vesc20}.

\begin{figure}[t]
\begin{center}
\includegraphics[width=1.00\columnwidth]{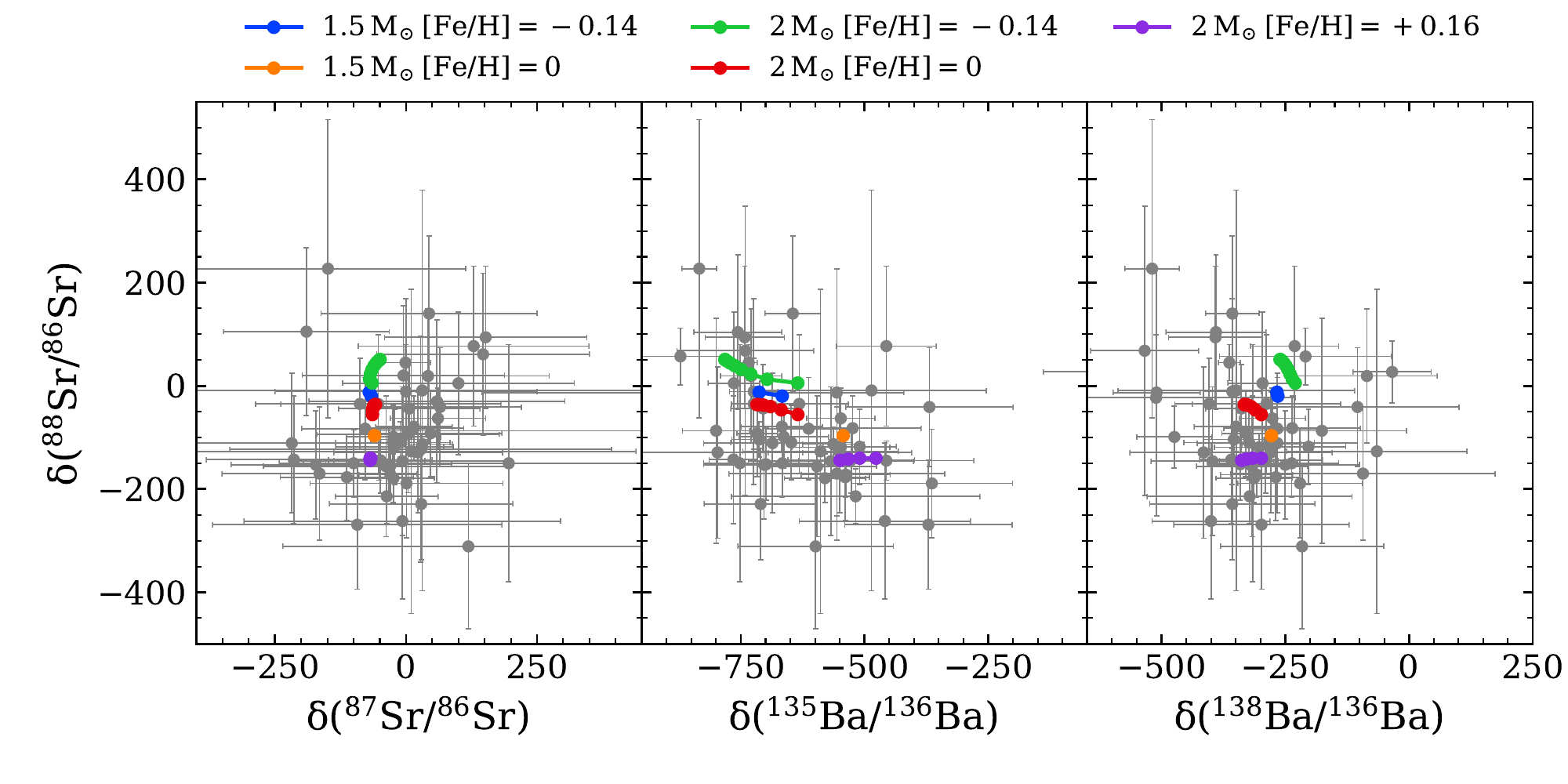}
\end{center}
\caption{\label{fig:grains_sr}Comparison between magnetic models predictions and $\delta$ values for isotopic ratios of Sr and Ba.
From left to right, ratios $^{88}{\rm Sr}/^{86}{\rm Sr}$ and $^{87}{\rm Sr}^{86}{\rm Sr}$, $^{88}{\rm Sr}/^{86}{\rm Sr}$ and $^{135}{\rm Ba}^{136}{\rm Ba}$, $^{138}{\rm Ba}/^{136}{\rm Ba}$ are represented. Data points refer to measurements from \cite{liu14b,liu15,step18}. The curves show the time evolution of the isotopic ratios in the stellar envelopes, with the dots along the lines representing the various TPs, corresponding to the C-rich phase.}
\end{figure} 
\section{Models at different metallicities}

We have computed 16 evolutionary models of 1.5$M_{\odot}$ and 2.0$M_{\odot}$ with different initial metallicities ($-$2.15 $\leq$ [Fe/H] $\leq$ +0.16).
We compared our \textit{s}-process nucleosynthesis calculations with the latest measurements of isotopic ratios of \textit{s}-elements in presolar SiC grains. We included data for Sr and Ba from \cite{liu14b,liu15,step18}. 
Figure \ref{fig:grains_sr} shows that new predictions for magnetic models with close-to-solar metallicity, are in general good agreement with $\delta$ (per mill) values for isotopic ratios of Sr and Ba, so confirming the results of \cite{palm18}.
Then, we compared the predictions for low and close-to solar-metallicity models with the spectroscopic data of post- and intrinsic C-rich AGB stars from \cite{reyn04,desm15,desm16,abia02,dela06,abia08,zamo09}.
\begin{figure}[t]
\begin{center}
\includegraphics[width=0.70\columnwidth]{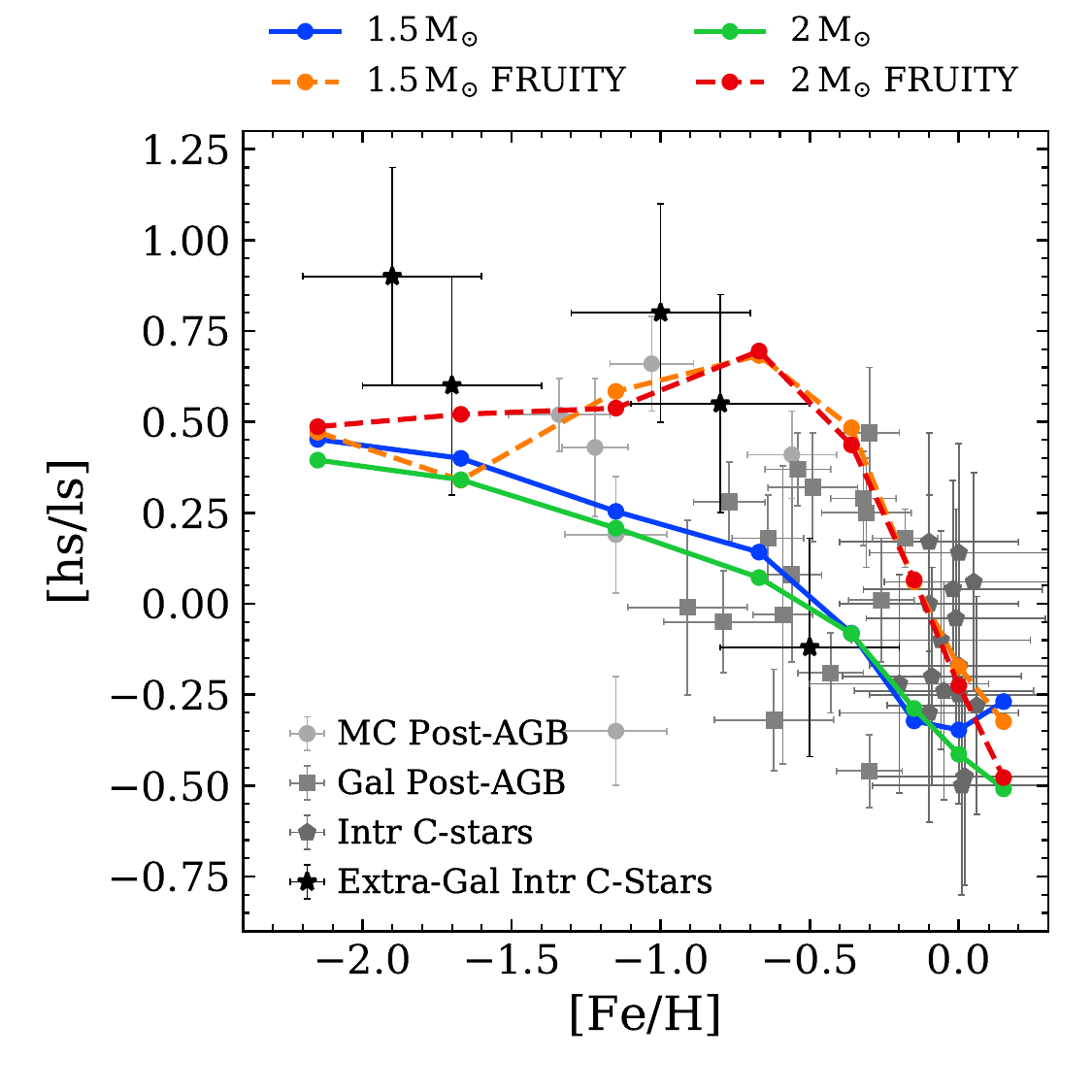}
\end{center}
\caption{\label{fig:postAGB_hsls_fe}Theoretical [hs/ls] as a function of the metallicity. Theoretical dashed curves refer to the non-rotating reference FRUITY models \cite{cris11}, while continuous curves refer to the magnetic models. Various symbols refer to: Magellanic Cloud post-AGB stars \cite{desm15}; galactic post-AGB stars \cite{reyn04,desm16}; galactic intrinsic C-rich stars \cite{abia02,zamo09}; extra-galactic intrinsic C-rich stars \cite{dela06,abia08}.}
\end{figure} 
Figure \ref{fig:postAGB_hsls_fe} shows that, while the reference FRUITY stellar models \cite{cris11} fail in reproducing the spectroscopic data, those models that take into account magnetic mixing for the formation of the $^{13}{\rm C}$-pocket agree on average with observations. In fact, while FRUITY models present a high [hs/ls] ratio over all the considered metallicity range, magnetic models show a low [hs/ls] ratio always but at a very low metallicity ([Fe/H] $\leq$ -1.5).
However, observational data show a large spread at a fixed metallicity that stellar models do not show. This is because we considered an average efficiency of magentic mixing. In principle, different stars, should exhibit a magnetic activity with different magnitude. As a consequence, the effect of stellar magnetic acitvity on the ongoing nucleosynthesis may differ from star to star. It is therefore reasonable to hypothesize that a spread in the magnetic mixing leads to a spread also in the s-process efficiency. This varibility could also be ascribed to the mass of the star itself, because the size of the magnetic pocket rapidly drops to very small values for higher-mass AGB stars \cite{vesc18}.

\section{Conclusions}

Most of what we know has been learned through a lengthy work with parameterized models, trying to constrain the parameters gradually, from the increasing accuracy of observations.
This recently allowed the development of physical models for the mixing
mechanisms required to produce the $^{13}{\rm C}$ neutron source.
Taking into account magnetic fields in radiative regions might be crucial in modeling the mixing episodes (e.g. through magnetic buoyancy).
First outcomes confirms the recent results from \cite{trip16,palm18}, showing that magnetic AGB models can reproduce the majority of isotopic ratios of mainstream grains. We find that the magnetic models set reproduces the dependancy of the [hs/ls] index, as a function of the metallicity, of post-AGB and intrinsic C-rich AGB stars. We speculate that the spread in the \textit{s}-process efficiency present in the observational data could be ascribed to a variable intensity of the magnetic activity of AGB stars.

\section*{References}
\bibliographystyle{iopart-num}
\bibliography{biblio}

\end{document}